\documentclass[%
 reprint,
 amsmath,amssymb,
 aps,longbibliography,
]{revtex4-1}

\usepackage{graphicx}% Include figure files
\usepackage{dcolumn}% Align table columns on decimal point
\usepackage{bm}% bold math
\usepackage{amssymb}
\usepackage{amsmath,bm}
\usepackage{tabularx}
\usepackage{rotating}
\usepackage{xcolor}
 \usepackage{siunitx}
 \usepackage[english]{babel}

\newcommand{\be}{\begin{equation}}
\newcommand{\ee}{\end{equation}}
\newcommand{\bee}{\begin{eqnarray}}
\newcommand{\eee}{\end{eqnarray}}

\begin{document}

%\preprint{APS/123-QED}

\title{Drag force in immersed granular materials}

\author{Tanvir Hossain}
\email{mhos0491@uni.sydney.edu.au}
\author{Pierre Rognon}%
 \email{pierre.rognon@sydney.edu.au}
\affiliation{Particles and Grains Laboratory, School of Civil Engineering, The University of Sydney, Sydney, NSW 2006, Australia}

\begin{abstract}
We investigate the drag forces acting on objects moving through a granular packing immersed in water. In this aim, we conducted uplift experiments involving pulling out horizontal plates at a prescribed velocity vertically. 
During these tests, we observed that the drag force reaches to peak at a low displacement and then decays. Results show that the peak drag force strongly increases with the velocity and depends on the plate size and grain diameter. We identify empirical scaling laws for these properties and introduce a Darcy-flow mechanism that can explain them. Furthermore, we conducted tests involving suddenly stopping the motion of the plate, which evidenced a progressive relaxation of the drag force in time. We discuss how a visco-elasto-plastic mechanical analogue can reproduce these dynamics. These results and analyses highlight fundamental differences in drag force between dry and immersed granular materials.
 %\pacs{45.05.+x}{45.70.-n}{45.50.-j }
%\keywords{Granular materials, drag forces}
\end{abstract}
\maketitle

\section{Introduction}
Drag forces are the resisting force acting on an object moving in a fluid. 
In Newtonian fluids, drag forces may be proportional to the object velocity $v$ at a low Reynolds number or proportional to $v^2$ at a high Reynolds number. They result from the action of viscous and inertial forces developing in the fluid being sheared and displaced around the object. Accordingly, this drag force vanishes when the object velocity tends to zero. Conversely, an object would continuously move at a non-null velocity when subjected to any non-null external force. 

Drag forces on objects embedded in dry granular materials are fundamentally different. These differences arise from the non-Newtonian behaviour of granular matter, which can deform elastically at low level of shear stress and flow plastically at high enough shear stress \cite{andreotti2013granular}. The first consequence of this behaviour is that objects embedded in granular packings may sustain a finite external force without continuously moving through the packing. The maximum force that the object can withstand while only inducing an elastic-like deformation of the packing is sometimes referred to as \textit{capacity} or \textit{peak drag}. In dry granular packings, this force is well understood. Many experiments and simulations showed that it is rate-independent at low object velocity. The drag force is then proportional to the hydrostatic stress and object surface area \cite{albert1999slow,albert2000jamming,albert2001granular,miller1996stress,nguyen1999properties,gravish2010force,costantino2011low,ding2011drag,guillard2013depth,takada2020drag}. This regime is called \textit{frictional drag} regime, by analogy with a Coulomb friction law. At high object velocity, the drag force increases quadratically with speed, which is reminiscent of a turbulent drag \cite{takada2020drag,percier2011lift,potiguar2013lift,takehara2010high,takehara2014high}. This regime is called \textit{inertial drag}. Like with Newtonian fluids, frictional and inertial granular drags result from the shear and the inertial displacement of the granular packing near the object.

The presence of water in granular materials is known to affect their rheological behaviour significantly. It can modify their effective viscosity and induce some delay in their deformation response to a stress change \cite{cassar2005submarine,rognon2010internal,rognon2011flowing,degiuli2015unified,ikeda2019universal}. This results from the additional dynamics associated with moving and shearing the fluid between grains in the pore network. 
Experiments consisting of moving a vertical rod at a constant velocity through an immersed granular materials recently revealed the linked between the rheology of immersed granular materials and a rate-dependent drag force developing on the rod \cite{allen2019effective}.  However, the effect of the presence of water on the peak drag force remains poorly understood.

In this Paper, we seek to establish how granular drag forces are affected by the presence of interstitial water. To this aim, we performed a series of elementary drag tests in fully immersed glass beads, varying the object size, its depth, its velocity and the grain size. The goal is to empirically establish the scaling of the drag force with these parameters and, subsequently, to infer the physical mechanisms at the origin of this behaviour. 

The Paper is organised as follows. Section \ref{sec:method} describes the experimental method used to measure drag forces. Section \ref{sec:maxdrag} presents the results of peak drag force and introduces an analytical model that captures its scalings. Finally, Section \ref{sec:relax} explores the drag force relaxation dynamics and introduces a mechanical analogue that helps explaining its origin.

\section{Experimental method}\label{sec:method}

\begin{figure*}[!t]
{\includegraphics[width=1\textwidth]{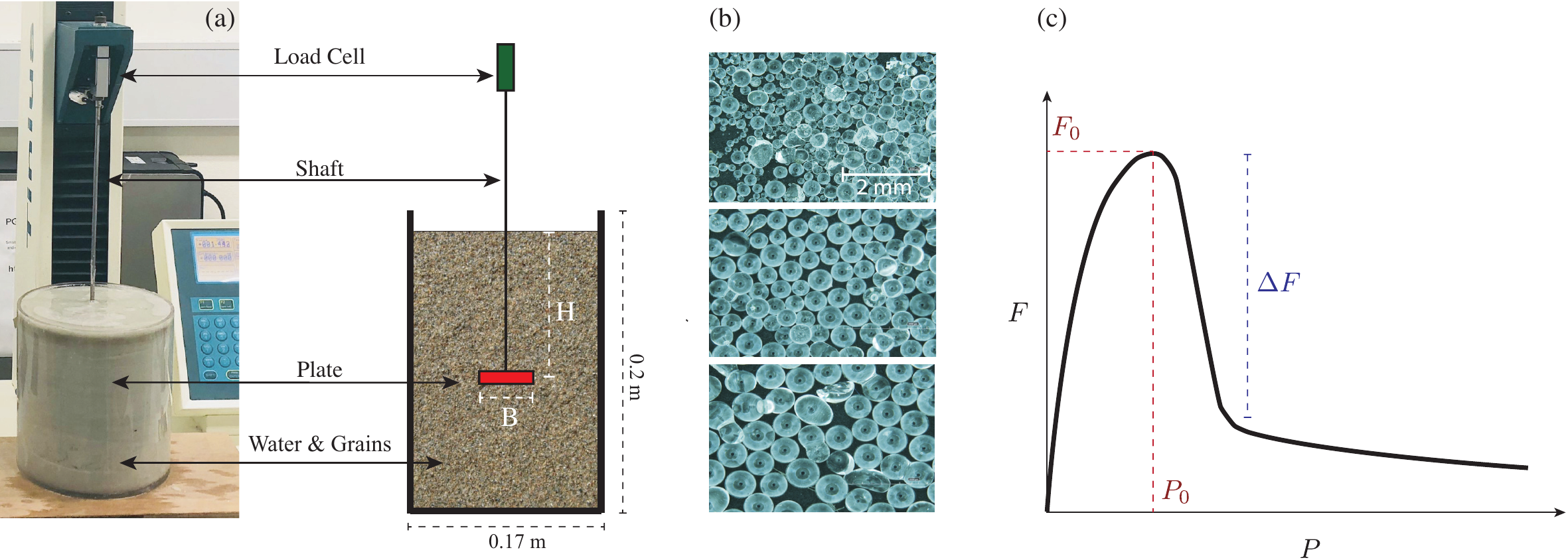}}
\caption{Experimental uplift tests. (a) Experimental set-up showing the loading frame, load cell, shaft, and the container filled with glass beads and water. (b) Photo of the glass beads used (Top:  $d$ = \SI{.3}{mm}; middle: $d$=\SI{.6}{mm};bottom: $d$= \SI{1}{mm}); the scale is the same on all three photos. (c) Illustration of a typical force (F) versus plate displacement (P) response obtained when pulling the plate vertically at a constant velocity.}
\label{fig1}       % Give a unique label
\end{figure*}

Uplift tests are performed using the experimental apparatus presented in figure \ref{fig1}a. This experimental set-up was previously used to measure drag forces in dry granular materials in Refs. \cite{dyson2014pull,hossain2020rate}. Tests involve placing a horizontal plate into a fully immersed packing of glass beads and pulling this plate vertically upward at a controlled velocity $v$. This velocity was reached by accelerating the plate at a constant acceleration during a fixed time period of \SI{0.01}{s}.
The plate is a PDMS cylinder of thickness \SI{4}{mm} and diameter $B$ ranging from \SI{30}{mm} to \SI{50}{mm}.  Grains are glass beads of diameter $d$ = \SI{.3}{mm}, \SI{.6}{mm} or \SI{1}{mm}, with a polydispersity of $\pm 10\%$. The density of the glass beads is $\rho_g = 2.6\times 10^3 $\SI{}{kg/m^3}. Grains and plate are fully immersed in the water; there are no air bubbles trapped into the packing. The granular packing is enclosed in a cylindrical plastic container of diameter \SI{170}{mm}. The plate is driven by a loading frame (H5KS Olsen Loading Frame) via a \SI{4} {mm} stainless-steel shaft and a force sensor. In the following tests, the plate motion is controlled by the loading frame via an electro-mechanical system comprised of a DC servo motor and a linear stage fitted with a screw.
The force required to achieve the prescribed plate motion is automatically adjusted and recorded during tests.

Preparation of the tests involves filling the container with a \SI{5}{cm} thick layer of grains and water, placing the plate at this location and gradually filling the rest of the container. The container is gently tapped while the mixture of grains and water are poured in order to produce dense packings. Visual inspection consistently revealed an absence of bubble near the container and no bubble coming up to the surface during uplift. This method produces packings with typical solid fraction  $\nu \approx 0.8$ and internal friction angle $\phi \approx 23^\circ$. The force sensor is zeroed just after the plate is placed at the desired location and before additional grains are poured above it. The force readings then directly measure the reaction force of the granular packing on the plate, excluding the weight of both the plate and shaft. This force corresponds to the drag force that is reported and analysed in the following.

We checked the repeatability of the experimental method by conducting several tests with similar conditions and measuring the maximum drag force $F_0$ as a point of comparison. Typically, we found a standard deviation lesser than $10\%$. We attribute these to variations in the granular packing configurations resulting from the pouring process, which may not be perfectly repeatable. We also checked for possible finite container size effect by repeating tests placing the largest diameter (\SI{50}{mm}) plate closer and closer to an edge, or closer and closer to the bottom. Unless the plate was placed very close to an edge (less than \SI{10}{mm}) or very close to the bottom (less than \SI{20}{mm}), we did not observe any significant variation in $F_0$. This suggests that tests performed in the centre of the container at a distance $\SI{50}{mm}$ from the bottom are not affected by the finite container size. 

\begin{table}[!t]
\begin{center}
\begin{tabular}{c|c|c|c|}
$d$ [mm]& $B$ [mm] & $H/B$ & $v$ [mm/s]\\
\hline
 0.3; 0.6; 1 & 30; 40;  50 & $3$ &  $1.6\times10^{-2}  \rightarrow 16$ 
 %$0.6  \rightarrow 1$ & $30\rightarrow 50$ & $3$ &  $1.6\times10^{-2}  \rightarrow 16$ 
 %$1.0  \rightarrow 1$ & $30\rightarrow 50$ & $3$ &  $1.6\times10^{-2}  \rightarrow 16$ 

%\hline  
\end{tabular}
\caption{Range of parameter explored experimentally, including grains size $d$, plate diameter $B$, plate embedment ratio $H/B$ and prescribed uplift velocity $v$.\label{tab:parameter}}
\end{center}
\end{table}%

\begin{figure}[!t]
{\includegraphics[width=1\columnwidth]{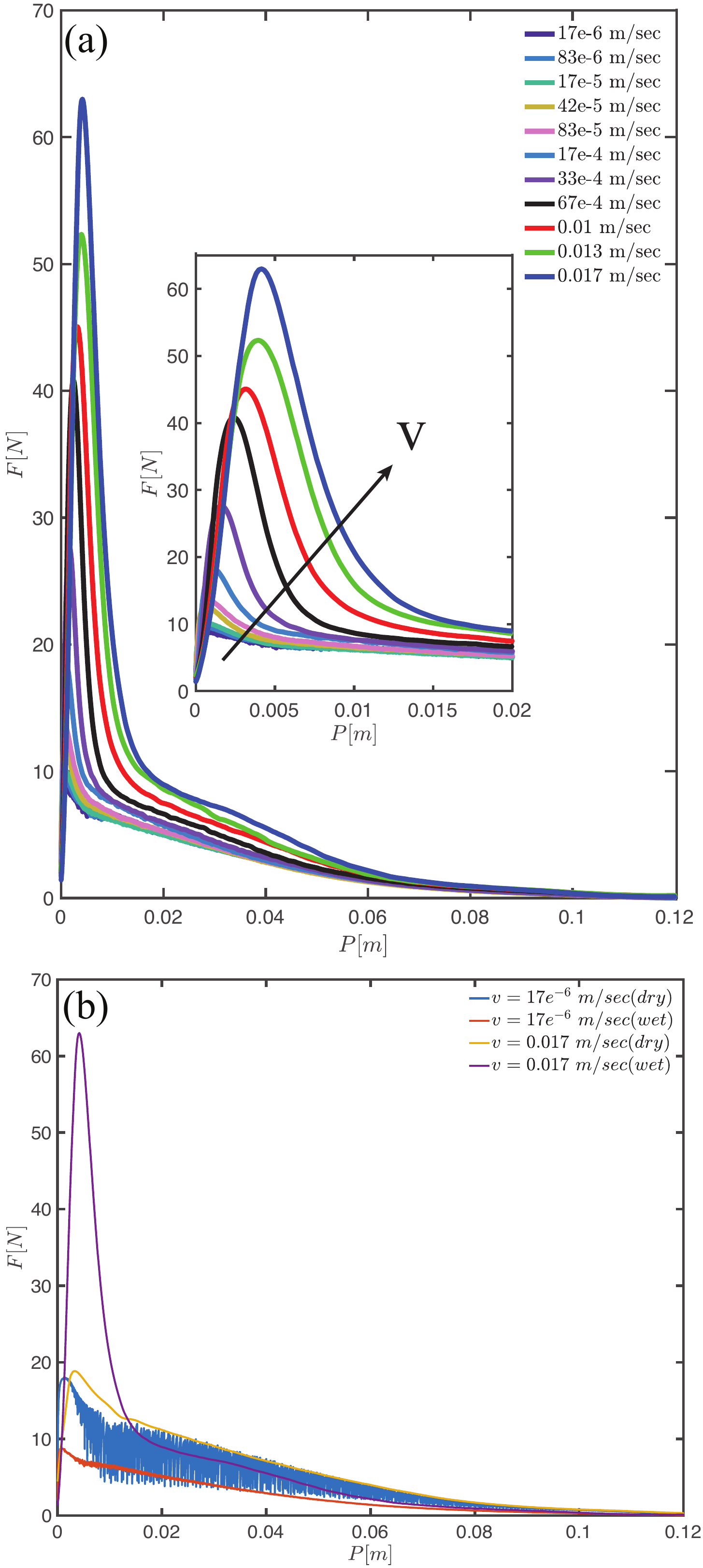}}
\caption{Drag force $F$ as a function of the plate displacement $P$ during uplift tests ($d=$\SI{.3}{mm}, $H=$\SI{120}{mm} and $B =$\SI{40}{mm}). (a) Tests performed in an immersed packing at different velocities; the inset focuses on small displacements. (b) Comparison between tests performed in a dry and in an immersed packing using the same plate and the same grains.
\label{fig2} }      % Give a unique label
\end{figure}

\begin{figure*}[!tb]
 {\includegraphics[width=1\textwidth,trim={0cm 0cm 0cm 0cm}, clip]{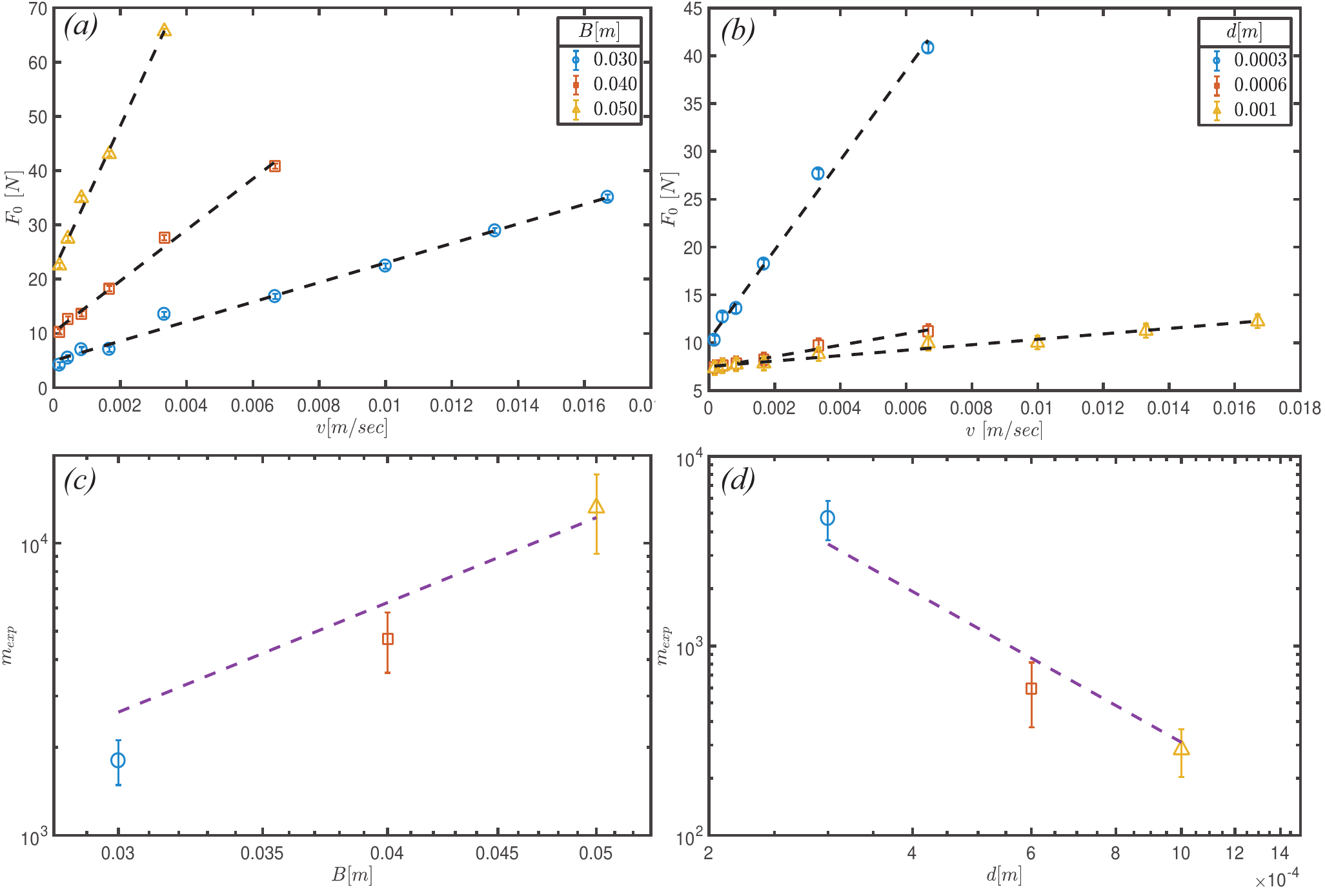}}
\caption{Effect of the uplift velocity $v$ on the peak drag force $F_0$. (a,b) Peak drag forces measured at different velocities with different grain sizes and plate diameters (see legends): (a) $d=$ \SI{0.3}{mm}, (b) $B=$ \SI{40}{mm}. Dashed lines correspond the best linear fits of the data by Eq. (\ref{eq:F_v}) using $F_0^{qs}$ and $m$ as fitting parameters. (c,d) Values of the parameter $m$ that best fit data in (a) and (b), respectively.  Error bars represent the confidence intervals for the parameter $m$ associated with the linear regression. Dashed lines show power laws with an exponent $3$ (c) and $-2$ (d) for a visual reference.  
}
\label{fig3}      
\end{figure*}

\section{Peak drag force}\label{sec:maxdrag}

This section focuses on the value of the peak drag force $F_0$ reached when uplifting the plate. It first presents the values of $F_0$ measured under different experimental conditions, as summarised in Table~\ref{tab:parameter}. It then introduces a physical model to rationalise these observations.

\subsection{Measurements}

The peak drag force $F_0$ was measured by performing uplift tests at a constant velocity with different plate size and grain size. Figure \ref{fig2} shows a set of drag force versus displacement curves obtained with $d=$\SI{0.3}{mm} and $B =$\SI{40}{mm} performed using immersed and dry grains. 
All curves are qualitatively similar to the illustration shown in figure \ref{fig1}c: the drag force first increases sharply to a peak $F_0$ and then decreases. 

With dry grains, the value of the peak force is similar for the fastest and the slowest velocities. At the slowest velocity, a drag instability develops after the peak is reached, leading to large drag force fluctuations, which does not develop with immersed grains - this behaviour is discussed in details in Ref. \cite{hossain2020rate}. Results show that the peak drag $F_0$ is strongly affected by the presence of water. It then significantly increases as the uplift velocity is increased. Figures \ref{fig3}a,b report the values of $F_0$ measured at different velocities using different grain sizes and with different plate diameters. These results suggest a increase for $F_0(v)$,  which we propose to decompose as follows:

\be \label{eq:F_v}
F_0 \approx F_0^{qs}+ m v
\ee

\noindent We define $F_0^{qs}$ as the peak drag in the quasi-static limit ($v\to0$). $m$ is a coefficient with a physical dimension of force per unit velocity. Figures  \ref{fig3}c,d indicate that this coefficient increases with the plate diameter, and decreases when using larger grain sizes. These dependencies are consistent with the following power laws: $m\propto B^3$ and $m\propto d^{-2}$. However, the available range of plate diameter and grain size only provides limited empirical support for the value of these exponents.

\begin{figure}[!t]
{\includegraphics[width=1\columnwidth]{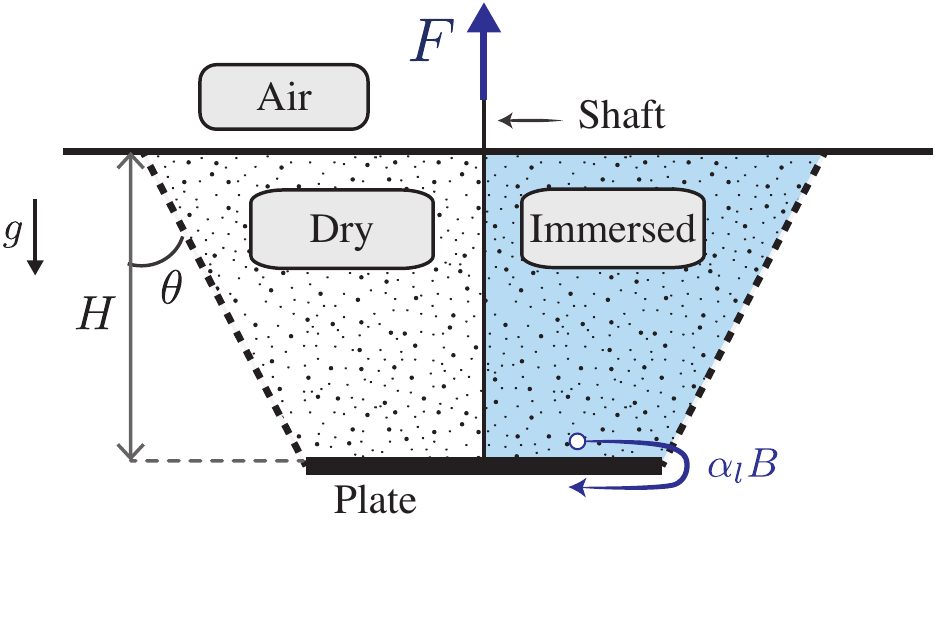}}
\caption{Origin of the peak drag force in dry (left) and immersed (right) granular packings.  In dry condition, $F_0$ corresponds to the weight of the grains comprised in the truncated cone (patterned area).
In fully immersed conditions, this weight accounting for buoyancy corresponds to the quasi-static component of $F_0$. We propose a Darcy-flow mechanism to explain the rate-dependent component of the drag force, whereby water recirculates around the plate through the granular matrix on a typical pathway length $\alpha_lB$. 
}
\label {fig4}       % Give a unique label
\end{figure}

\subsection{Physical origin}

To understand the physical origin of the peak drag force scaling observed in the previous Section, let us first consider similar uplift experiments conducted using dry grains such as those presented in Ref. \cite{hossain2020rate}. Without water, the peak drag force $F^{dry}_0$ was found to be rate-independent in the range of uplift velocities considered here. The consensus is that $F^{dry}_0$ corresponds to the weight of a volume of grains located above the plate, and being supported by the plate. The precise shape of this volume depends on the internal friction coefficient of the packing, \cite{das2013earth,meyerhof1968ultimate,rowe_behaviour_1982,murray_uplift_1987,merifield2006ultimate,kumar2008vertical},  the object shape \cite{dyson2014pull,khatri2011effect,bhattacharya2014pullout,askari2016intrusion,giampa2018effect} and the grain size \cite{sakai1998particle,athani2017grain,costantino_starting_2008}. For a circular plate, this volume is illustrated on figure \ref{fig4}: it is a truncated cone originating from the plate and expending toward the free surface with an angle $\theta \approx \phi$. The corresponding model for a circular plate of area $S=\pi B^2/4$ is:

\bee \label{eq:F0dry}
F_0^{dry} &=& \nu \rho_g g H S f(H,B)\\
f(H,B)&=&\frac{1}{3} \left[\left(1+2\frac{H}{B} \tan \theta \right)^2 +2 \frac{H}{B} \tan \theta + 2 \right]. 
\eee

\noindent $\nu \rho_g g H S$  corresponds to the weight of the cylinder of grains located above the plate.  The function $f$, which is greater than one, accounts for the truncated cone shape; $f=1$ would correspond to a cylinder shape.

Based on this mechanism, we propose that drag forces in immersed packings involve two components corresponding to two distinct physical processes:

\be \label{eq:F0decomposition}
F_0 = F_0^{qs} + F^{vis}
\ee

\noindent We assume that $F_0^{qs}$ is a rate-independent component resulting from the weight of the truncated cone. However, the presence of water induces buoyancy, which reduces the effective weight of this truncated cone. Accordingly, we propose to express the quasi-static component as: 

\bee
F_0^{qs} &=& \nu (\rho_g-\rho_w)  g H S f(H,B) \\
&=&  F_0^{dry} \left( 1- \frac{\rho_w}{\rho_g}\right) \label{eq:F0qs}
\eee

\noindent where $\rho_g$ is the density of glass and $\rho_w$ the density of water. Table \ref{tab:comparison} shows that the prediction of this  model approximately matches the quasi-static maximum drag $F^{qs}_0$ obtained by fitting our experimental data $F_0(v)$ with Eq. (\ref{eq:F_v}), with both $F^{qs}_0$ and $m$ as free fitting parameters. 
This indicates that the \textit{truncated cone} mechanism is still relevant in the presence of water and that it is at the origin of the quasi-static component of the maximum drag force. 

\begin{table}[]
\begin{center}
\begin{tabular}{c|c|c|c|cl}
%\hline
 $B$ [mm]& $d$ [mm] & $H/B$ & Predicted $F^{qs}_0$, [N] & Measured $F^{qs}_0$, [N]\\
\hline
 $30$ & $0.3$ & $3$ &  $4.3$ & $5.0$\\
\hline
 $40$ & $0.3$ & $3$ &  $10.3$ & $10.3$\\
\hline
 $50$ & $0.3$ & $3$ &  $20.2$ & $21.8$\\
\hline
 $40$ & $0.6$ & $3$ &  $10.3$ & $7.4$\\
\hline
 $40$ & $1$ & $3$ &  $10.3$ & $7.5$\\
\hline  
\end{tabular}
\caption{Quasi-static peak drag force $F^{qs}_0$: prediction of the proposed model (Eq. (\ref{eq:F0qs}) using $\theta = \phi$, $\nu=0.8$ and $\rho_w=10^3$\SI{}{kg/m^3}) and measurements (obtained by fitting of the experimental data $F_0(v)$ with Eq. \ref{eq:F_v}). \label{tab:comparison}}
\end{center}
\end{table}%

$F^{vis}$ is a rate-dependent component of the peak drag force. According to our experimental observations, $F^{vis}$ should be proportional to the uplift velocity: $F^{vis} = m v$. In Ref. \cite{athani2019inertial}, a numerical study of uplift in dry grains evidenced a similar linear increase in peak drag force, which attributed to the initial acceleration of the plate and the inertial resistance of the packing. The peak drag force was then reached when the plate stopped accelerating. 
In our experimental conditions, the plate is accelerated during a fixed period of time of \SI{0.01}{s}, corresponding to an acceleration ranging from $1.6\times 10^-{4}$ g to $0.16$ g. The plate stops accelerating at small displacements lesser than \SI{0.16}{mm}. Figure \ref{fig2}a shows that the drag force keeps increasing during its subsequent steady motion. Figure \ref{fig2}b further shows that, without water, the peak drag force is rate independent with these experimental conditions. This indicates that another mechanism that the plate acceleration causes the rate-dependent peak drag force. We attribute this component to the displacement of water induced by the plate's motion, and propose the following physical mechanism to explain it.

We consider that the plate's motion induces some flow of water from above the plate to below the plate. At small displacements ($P\leqslant P_0$), we assume that the granular matrix has not significantly deformed plastically. The flow of water thus corresponds to a Darcy flow through an immobile porous matrix. While this flow is driven by complex local gradients of pore pressure, we propose to model in a simplified manner to estimate its contribution to the drag force. 
We consider that the flow is driven by pressure drop $\Delta p$ corresponding to the difference in pore water pressure above and below the plate. The net force acting on the plate is related to this pore water pressure difference: 

\be \label{eq:FvisDP}
F^{vis} = \Delta p \pi B^2/4.
\ee 

\noindent We further relate the pressure drop to the plate velocity by introducing Darcy's law, involving the permeability of the packing $K$ and an effective water pathway $\alpha_l B$: 

\be \label{eq:Darcy}
v = \frac{K}{\eta} \frac{|\Delta p|}{\alpha_l B}.
\ee

\noindent where $\eta$ is the viscosity of water. This formulation considers an effective pathway of the water around the plate given by $\alpha_l B$: it is proportional to the plate size via a dimensionless constant $\alpha_l$, which value is to be determined. This pathway is illustrated in figure \ref{fig4}. Finally, we use the Carman-Kozeny model to relate the granular packing permeability $k$ to the typical pore cross-section are, or to the grain size \cite{carman1939permeability,rognon2014explaining}:

\be \label{eq:KC} 
K= \alpha_K \frac {(1-\nu)^{3}}{\nu ^{2}} d^{2} 
\ee

\noindent where $\alpha_K$ is a dimensionless constant that depends on the grain shape and polydispersity. For perfectly spherical  and mono-disperse grains, it is of the order of $\alpha_K\approx 1/150$. 

Combining Eqs. (\ref{eq:FvisDP}), (\ref{eq:Darcy}) and (\ref{eq:KC}) leads to the following expression for the viscous component of the peak drag force:

\bee \label{eq:Fvis}
F^{vis} &=& m v \\
m &=& \alpha \eta \frac{B^3}{d^2} \\
 \alpha &=& \frac{\pi}{4}  \frac{\alpha_l}{\alpha_k} \frac{\nu^2}{(1-\nu)^3}
\eee

This model predicts that the viscous component of the peak drag force is proportional to the plate velocity. Furthermore, it highlights an expression for the coefficient $m$; using a porosity of $\nu=0.8$ and $\alpha_k=1/150$, the prediction for the coefficient $m$ becomes:

\bee 
m &=& 150 \frac{\pi}{4}  \alpha_l \frac{.8^2}{(.2)^3} \eta \frac{B^3}{d^2}\\
      &\approx& 9\,425 \times\alpha_l  \eta \frac{B^3}{d^2} \label{eq:mth}
\eee

%\SI{}{Pa.s}

\noindent In our experimental conditions, we estimate that the viscosity of the water is $\eta = 8.9\times 10^{-4}$ $Pa.s$. The only remaining unknown to estimate $m$ is then the value of the coefficient $\alpha_l$, which reflects the effective pathway length of the liquid in unit $B$. To determine it, we fitted all the values of $m$ that we obtained experimentally with Eq.(\ref{eq:mth}) using $\alpha_l$ as a sole free fitting parameter. The best fit was obtained with $\alpha_l \approx 0.62$. This indicates that the effective pathway for the Darcy flow is slightly larger than half the plate diameter, which is consistent with the mechanism of recirculation around the plate illustrated in figure \ref{fig4}.

\begin{figure}[t!]
 {\includegraphics[width=1\columnwidth]{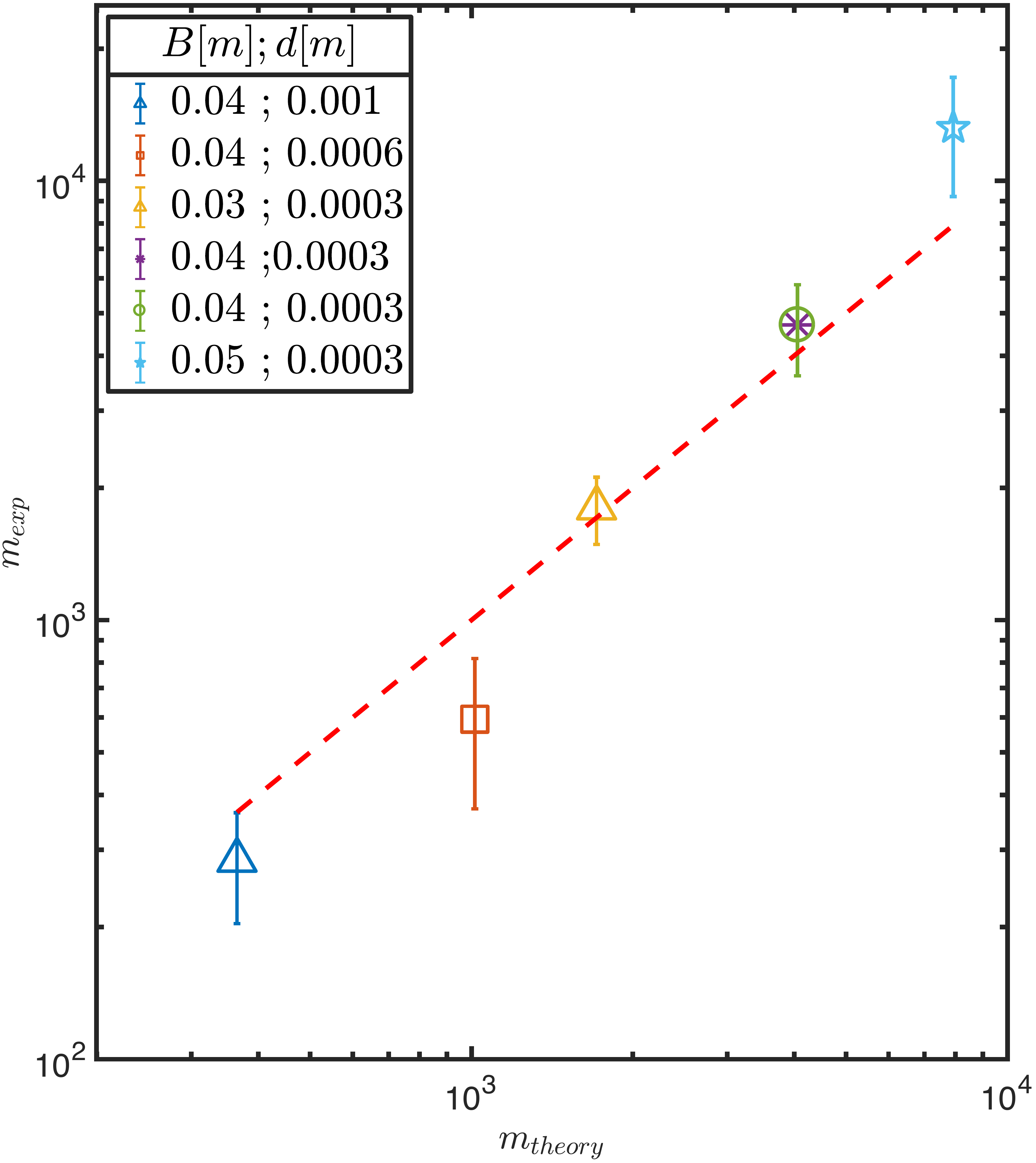}}
\caption{Comparison of the parameter $m_{exp}$ (measured experimentally by fitting the data $F_0(v)$  by Eq. (\ref{eq:F_v})) with the value $m_{theory} $ (predicted by the proposed model in Eq. (\ref{eq:mth}) using $\alpha_l=0.62$). The dashed line represents the function $m_{exp}=m_{theory}$ for visual reference.
}
\label{fig5}      
\end{figure}

Figure \ref{fig5} compares the prediction of this model by plotting Eq. (\ref{eq:mth}) versus the experimentally determined values of $m$ for different grain sizes and plate diameters. The near match supports the credibility of the proposed Darcy flow mechanism as a cause for the viscous component of the peak drag force. 

\begin{figure*}[htb!]
 {\includegraphics[width=1\textwidth]{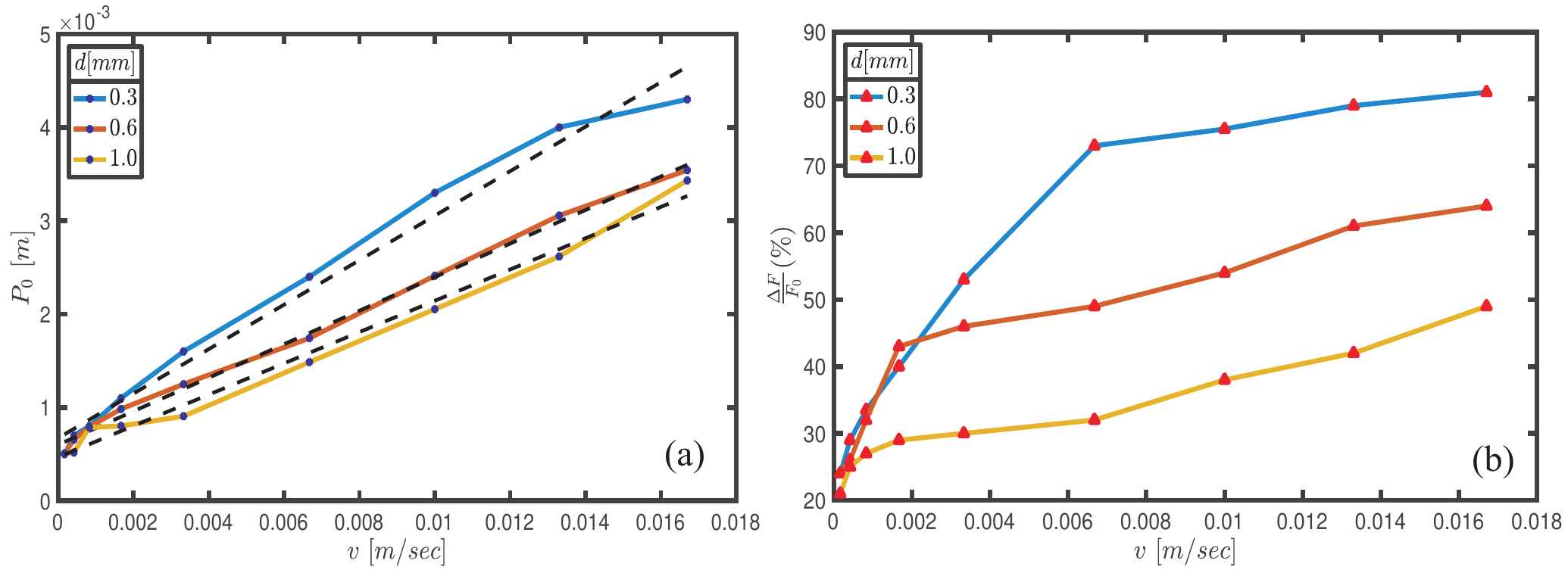}}
\caption{ Post-peak behaviour measured from data shown on figure \ref{fig2} ($B=$\SI{40}{mm}). (a) Displacement of the plate $P_0$ when the peak drag force is reached; 
The dashed line represents the best linear fit using Eq. (\ref{eq:P0}), obtained with $P_0^{qs}\approx $ 0.67; 0.6; 0.47 mm and $\lambda_0 \approx $ 0.24; 0.18; 0.17 s for grain size of 0.3; 0.6; 1 mm, respectively.
(b)  post peak relative drag force drop $\Delta F/F_0$ (see figure \ref{fig1}c).}
\label{fig6}      
\end{figure*}

\subsection{Post-peak behaviour}

In addition to the velocity-driven increase in peak drag force, the drag force versus displacement curves shown in figure \ref{fig1} highlight two features.

Firstly, the peak drag force is reached at a displacement $P_0$, which increases when the velocity is increased. Figure \ref{fig6}a reports the values of $P_0(v)$ measured from the data shown in figure \ref{fig2}. As a first approximation, it suggests a linear increase which we propose to decompose as follows:

\be \label{eq:P0}
P_0 \approx P_0^{qs} + \lambda_0 v
\ee

\noindent $P_0^{qs}$ is the quasi static peak displacement, corresponding to the limit $v\to 0$.  $\lambda_0$ is a characteristic time which measures how long the drag force takes to build up to its peak. The best linear fit of the peak displacement $P_0(v)$ are obtained using $P_0^{qs} \approx $ \SI{0.58}{mm} and $\lambda_0 \approx $ \SI{0.25}{s}.
Interestingly, the linearity of $P_0(v)$ indicates that the peak drag force is not simply reached when a fixed displacement is reached. It rather suggests that it takes a finite amount of time $\lambda_0$ to reach it.

Secondly, figure \ref{fig1} shows that the drag force sharply decreases after the peak displacement is reached ($P>P_0$). We measured the post-peak drop in drag force by the quantity:

\be
\Delta F = F_0 - F_{inflection}
\ee 

\noindent where $F_{inflection}$ is the drag force measured at the inflexion point of $F(P)$ following the peak displacement (see figure \ref{fig1}c). 
Figure \ref{fig6}b shows that the drag force drop $\Delta F$ can reach up to $80\%$ of the peak force $F_0$ at high uplift velocities. This suggests that the viscous component of the peak drag force $F^{vis}$ virtually vanishes after the peak is reached.  

While the Darcy flow mechanism can explain the velocity dependence of the peak drag force, it is not sufficient to explain these two observations.
The next section will seek to identify the relevant missing physical mechanisms. 

\begin{figure}[htb!]
 \includegraphics[width=1\columnwidth]{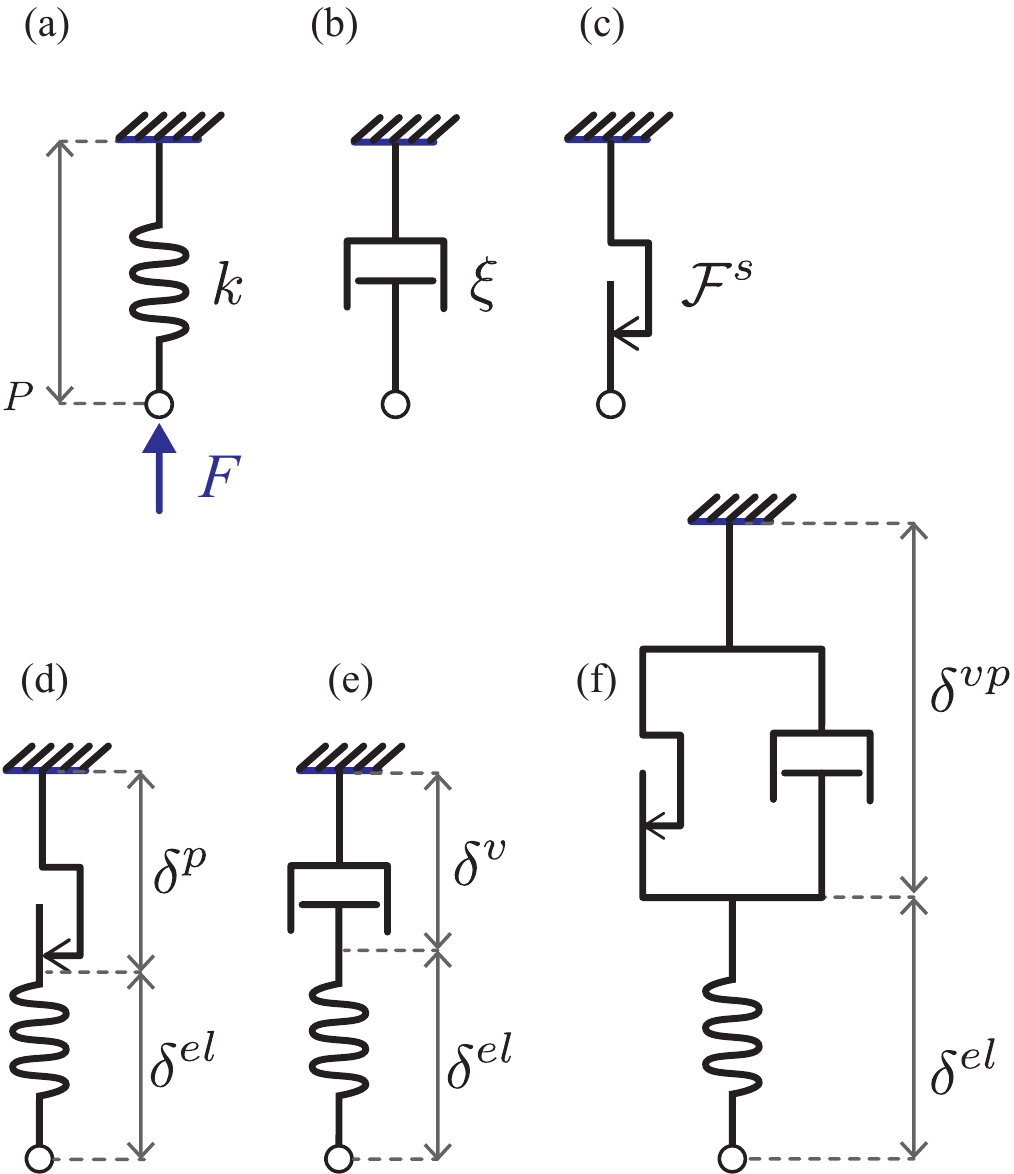}
\caption{Examples of mechanical analogue: (a) spring, (b) dashpot and (c) slider elements; (d) elasto-plastic, (e) visco-plastic and (f) elasto-visco-plastic analogues.}
\label{fig7}      
\end{figure}

\begin{figure*}[htb!]
 {\includegraphics[width=1\textwidth] {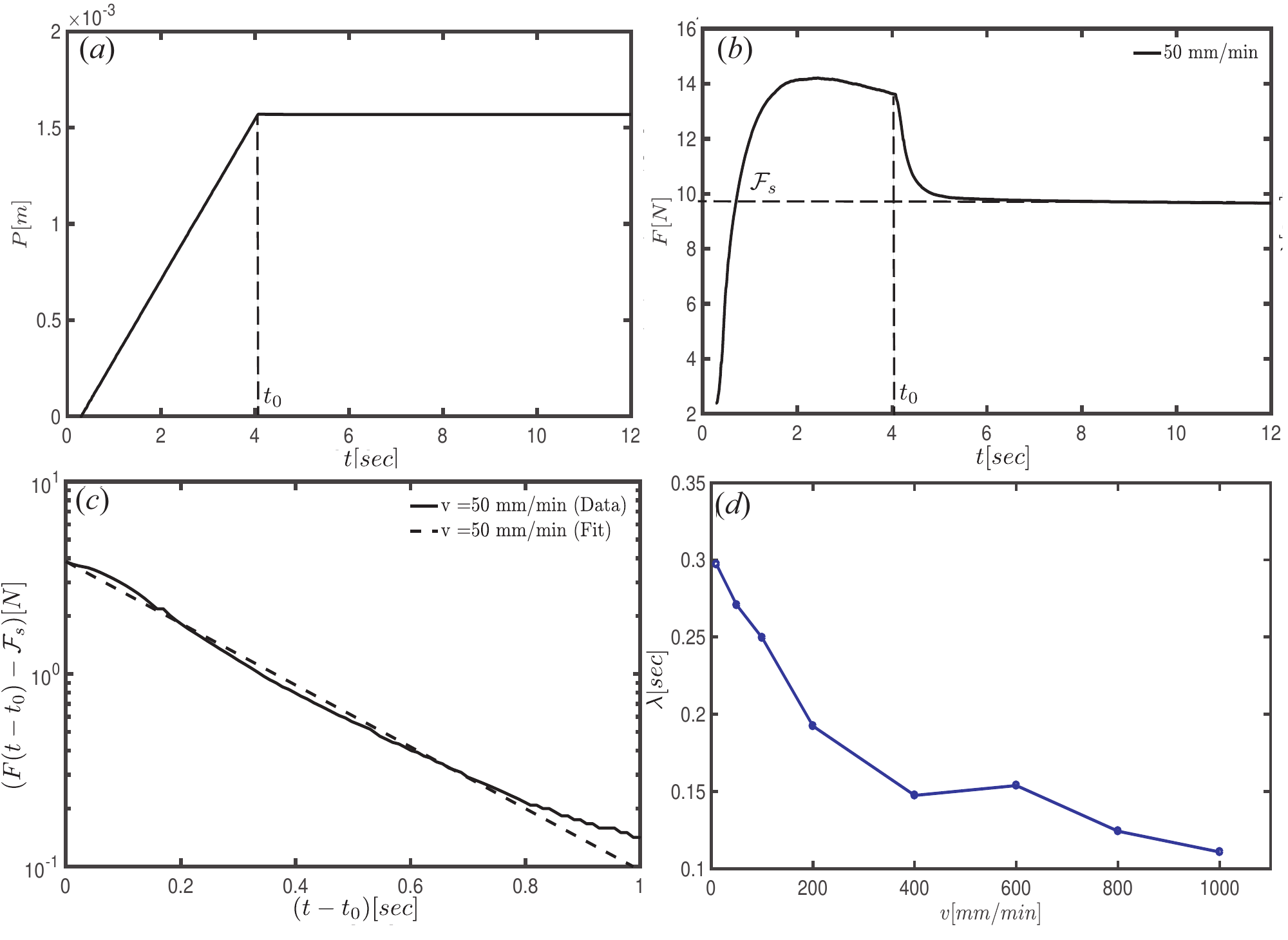}}
\caption{Relaxation tests performed in \SI{0.3}{mm} grains. (a,b) Displacement $P$ and measured drag force during a test where the plate (B=\SI{40}{mm}) is uplifted at a constant velocity $v=$\SI{.4}{mm/s} during $t_0$=\SI{4}{s} and then stopped.  (c) Semi-log plot of the drag force evolution during relaxation (same data as in (b)); The dashed line represent the best fit using  the exponential decay in Eq. (\ref{eq:relax}), using $\lambda$ as sole fitting parameter. (d) Values of relaxation time $\lambda$ obtained by similarly fitting tests conducted with different velocities $v$. 
}
\label{fig8}      
\end{figure*}
\section{Drag force mechanical analogue}\label{sec:relax}

The previous section has highlighted three main properties of the drag force in fully immersed packings: (i) the peak drag force $F_0$ increases linearly with the uplift velocity (ii) the peak displacement $P_0$ also increases linearly with the uplift velocity and (iii) the drag force strongly drops after the peak is reached. In this section, we seek to develop a simple mechanical analogue in order to understand the origin of these properties better. 

\subsection{Elementary mechanical analogues}

We consider the three elementary mechanical analogues that are illustrated in figures \ref{fig7}a-c: a linear spring, a linear dashpot and a slider. 
By analogy with our drag experiments, we call $P$ the distance between the moving bottom of the analogue and its fixed top, $\dot P = v$ the upward velocity and $F$ the force applied at point $P$ to generate this motion. The mechanical behaviour of these three analogues is given by $F = k P$ for the spring where $k$ is a spring stiffness parameter, $F = \xi \dot P$ for the dashpot where $\xi$ is the viscous parameter, and $\dot P = 0$ if $F<\mathcal{F}^s$ and $F=\mathcal{F}^s$ otherwise for the slider (which \textit{sticks} or \textit{slips}, respectively) where $\mathcal{F}^s$ is a force threshold parameter.

 Combining these elements in parallel makes them experience the same displacement. Combining them in series make them experience the same force. 
 Such combinations produce a variety of mechanical behaviours. For instance, the simple elasto-plastic analogue presented on figure \ref{fig7}d would be sufficient to predict a linear increase in drag force $F= kP$ up until a maximum value $F=\mathcal{F}^s$, and a maximum displacement $P_0 = \mathcal{F}^s/k$.
Considering that the slider parameter is the peak drag force, $\mathcal{F}^s=F_0$, this analogue would capture the evolution of the drag force in dry conditions until $P=P_0$. Then, the parameter $k$ would account for an effective stiffness of the granular packing, and the slider would account for its plastic deformation.
However, this elasto-plastic analogue is rate-independent. It is therefore not sufficient to capture the drag force in immersed packings.

The Maxwell visco-elastic analogue illustrated in figure \ref{fig7}e is rate-dependent. Compressed at a constant velocity $\dot P=v$, the force would reach a steady state $F = \xi v$ after gradually increasing during a characteristic time $ \xi/k$. However, it predicts a maximum drag force of $\xi v$, which does not completely match the observed linear increase $F_0(v)$, given that it includes a non-null peak force $F_0^{qs}$ for $v\to0$

The elasto-visco-plastic analogue illustrated on figure \ref{fig7}f can capture this non-null peak force in the quasi-static limit by introducing a slider with $\mathcal{F}_s=F_0^{qs}$. Under constant velocity, this analogue responds with a linear increase of the force $F(t) = k P$ until $F_0^{qs}$ is reached. Then the slider moves freely, and the force in the dashpot increases from $0$ to a value of $v\xi$ after a characteristic time $\xi/k$. This results in a maximum drag force of $F_0 = F_{0}^{qs} + \xi v$, which is consistent with our experimental observation.
Furthermore, this analogue predicts that the maximum drag force would be achieved at a displacement of the order of  $P_0 \approx \frac{F_0^{qs}}{k} + \frac{\xi}{k} v$, which is also consistent with the observed linear increase for $P_0(v)$.

\subsection{Drag force visco-elastic relaxation}

The simple elasto-visco-plastic analogue on figure \ref{fig7}f qualitatively captures the behaviour of the drag force before the peak is reached as well as its peak value. It further predicts that, if the motion of the plate is stopped after the peak is reached ($\dot P = 0$), the drag force should gradually relax in time toward the value of the slider force threshold $\mathcal{F}_s$ with a characteristic time $\xi/k$.

To verify whether this relaxation actually occurs, we conducted a series of tests whereby the plate is uplifted at a constant velocity $v$ until the peak force is reached, and then stopped. These tests were conducted at different velocities, using a plate of diameter \SI{40}{mm} and a grain size of \SI{0.3}{mm}.
An example of the plate displacement and drag force evolution of such tests is shown on figures \ref{fig8}a,b. It confirms that the drag force relaxes once the plate motion is stopped. Furthermore, figure \ref{fig8}c indicates that this relaxation is exponential, which also matches the analogue dynamics:

\be \label{eq:relax}
F(t-t_0)-F(t_0) \approx \left( \mathcal{F}_s - F(t_0)\right) e^{-\frac{t-t_0}{\lambda}}
\ee

\noindent $t_0$ is the time at which the plate motion was stopped and $\lambda$ is a characteristic time scale of the relaxation. $\mathcal{F}_s$ corresponds to the drag force at time $t\to\infty$. In this experiment, we found $\mathcal{F}_s =$ \SI{9.8}{N}, which is close to, but slightly lower than the measured quasi-static drag force: $F_0^{qs} =$ \SI{10.3}{N}.

We measured the relaxation time $\lambda$ on tests performed at different initial velocities by fitting the drag force relaxation measurements with Eq. (\ref{eq:relax}) using $\lambda$ as sole fitting parameter. Both $\mathcal{F}_s =F(t \to \infty)$ and $F(t_0)$ are directly measured. Results shown in figure \ref{fig8}d indicate a slight decrease of $\lambda(v)$, with values ranging from \SI{0.3}{s} to \SI{0.11}{s} in the explored range of velocity. 
We have conducted similar relaxation tests varying the time $t_0$ at which the plate is stopped, choosing values shortly before and after the peak force is reached. All tests led to a similar relaxation dynamics with no significant influence on the time constant $\lambda$. 
Interestingly, these values are close to the time $\lambda_0\approx $  \SI{0.24}{s} needed for the viscous drag force to build up. This suggests that both viscous drag force build-up and relaxation could result from a similar visco-elastic dynamics.  

However, we could not find a set of spring constant $k$ and viscous parameter $\xi$ that would lead to simultaneously match the measurements of (i) the increase in drag force at small displacements, (ii) the peak force $F_0$ and (iii) the relaxation time $\lambda$.
According the mechanical analogue of figure \ref{fig7}d, these are given by: $F(P) = kP$ for $P<\mathcal{F}_s/k$, $F_0 = \mathcal{F}_s + \xi v$ and $\lambda = \xi/k$. This means that prescribing a stiffness $k$ and a viscous parameter $\xi$ in order to match the pre-peak and peak drag force $F_0$ determines the relaxation time.

\begin{figure}[tb!]
 \includegraphics[width=0.6\columnwidth]{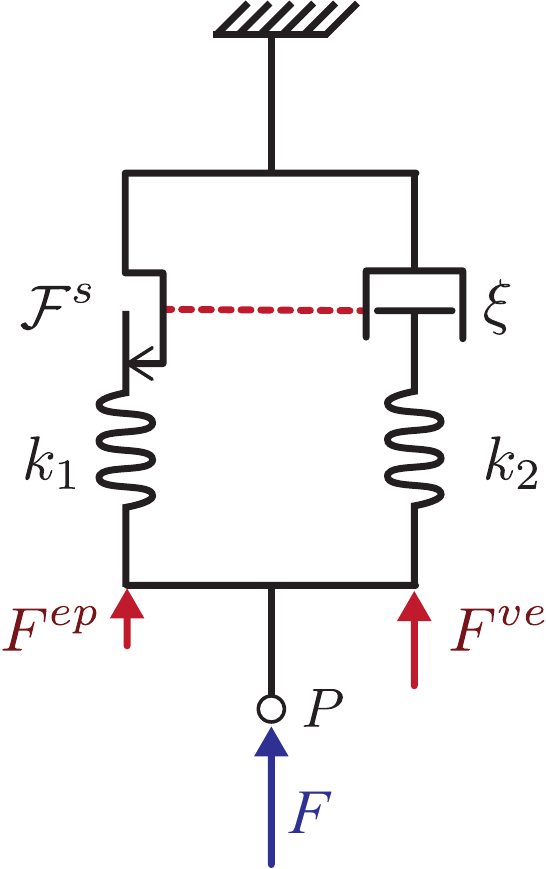}
\caption{Proposed mechanical analogue for the drag force in immersed granular packings. The force $F$ is distributed between the elasto-plastic element (left) and the visco-elastic element (right): $F = F^{ep}+F^{ve}$. Both elements undergo the same deformation $P$ and deformation rate $\dot P$ at any point in time. The dashed red line indicates that large slider deformations may lead to lowering the dashpot viscous parameter $\xi$; this could explain the post-peak drop in drag force observed on figure \ref{fig2}.}
\label{fig9}      
\end{figure}

%\begin{figure*}[htb!]
 %{\includegraphics[width=1\textwidth]{figure/spring_pore.eps}}
%\caption{Relaxation test results (a) the force-displacement plot with corresponding fit at three uplift speeds, i.e., $10$ $mm/min$, $400$ $mm/min$,$1000$ $mm/min$ .(b) spring constants, $k_1$ obtained from the slope of the fitted lines of (a) are plotted with %uplift speeds ranging from $10 \sim 1000$ $mm/min$. It is observed that the spring constant decreases as the uplift speed increases . (c)  pore water pressure vs uplift speeds curve. It is observed that the pore water pressure increases as the uplift speed %increases. and (d) maximum uplift capacity vs uplift speed curve. It is observed that the uplift capacity increases as the uplift speed increases.}
%\label{model_relax}      
%\end{figure*}

\subsection{Proposed drag force analogue}

\begin{figure*}[htb!]
 {\includegraphics[width=1\textwidth]{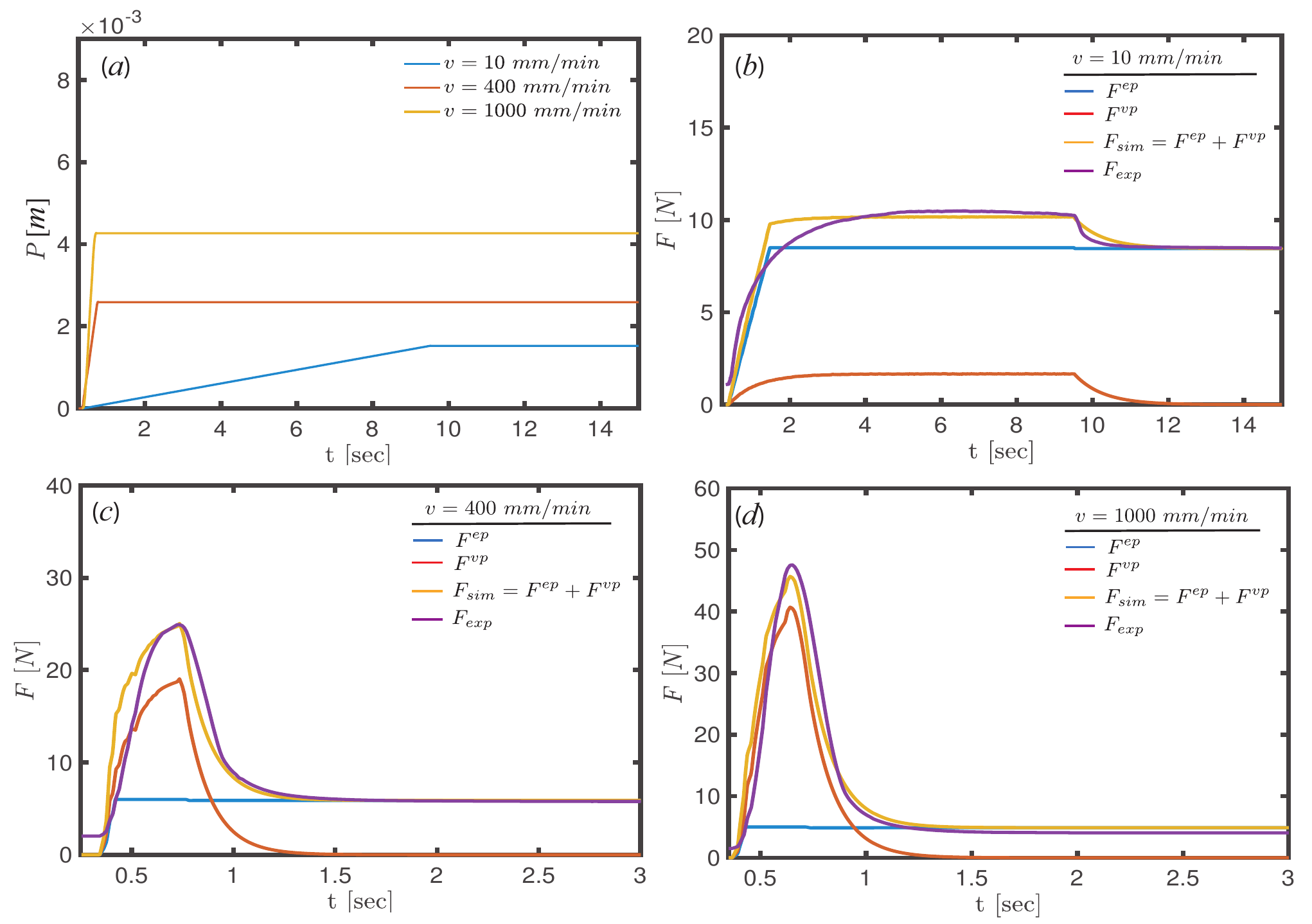}}
\caption{Drag force relaxation dynamics. (a) Three experimental plate displacement where a constant velocity $v$ is applied before the plate is stopped. (b-d) drag force evolution corresponding to tests on (a): v = \SI{10}{mm/min} (b), \SI{400}{mm/min} (c) and \SI{1000}{mm/min} (d). (b-d) compare the experimental drag force measurements ($F_{exp}$)  to the predictions of the proposed visco-elasto-plastic model on figure \ref{fig9} ($F_{sim}$, together with its elasto-plastic and visco-elastic components). The model parameter used are summarised on Table \ref{tab:model_para}.
}
\label{fig10}      
\end{figure*}

We propose to resolve this discrepancy by introducing a second spring element in the visco-elasto-plastic model, as illustrated on figure \ref{fig9}. The resulting analogue is comprised of an elasto-plastic element and a Maxwell visco-elastic element connected in parallel. 
A possible interpretation for these two springs would be that that the elasto-plastic spring corresponds to the elastic deformation of the matrix under the action of a force originating from the plate and carried via contacts, while the visco-elastic spring corresponds to the deformation of the granular matrix under the action of the Darcy flow, which can be seen as the action of the forces carried by the water. The analogue mechanical response to a relaxation test is:

\bee
F(P) &=& (k_1+k_2) P \text{ for } t \to 0  \label{eq:elastic} \\
F_0 &=& \mathcal{F}_s+ \xi v \label{eq:peakmodel}\\
\lambda &=& \xi/k_2 \label{eq:relaxmodel}\
\eee

Eq. (\ref{eq:elastic}) corresponds to the instantaneous elastic response at the beginning of the test, when the slider sticks and does not deform while the dashpot had no time to deform. Eq. (\ref{eq:peakmodel}) represents the maximum force achieved under a constant velocity $\dot P= v$. Eq. (\ref{eq:relaxmodel}) represents the characteristic relaxation time of the drag force occurring when the plate is stopped.

In order to compare the response of this analogue to our experimental results, we have numerically resolved it by integrating the following system of equation:

\bee
F &=& F^{ep}+F^{ve}\\
   \dot F^{ep}&=& 
\begin{cases}
   k_1 \dot P \text { if } F^{ep}<\mathcal{F}^s\\
    0    \text{ otherwise}
\end{cases} \label{eq:Fep}\\
\dot{P} &=& \frac{ \dot F^{ve}}{k_2} +\frac{ F^{ve}}{\xi} \label{eq:Fve}\\
P(t=0) &=& 0; \dot P(t<t_0) = v; \dot P(t\geqslant t_0) = 0;\\
 F^{ep}(t=0)&=& F^{ve}(t=0) =0
\eee

\noindent The numerical resolution consists of solving for the elasto-plastic and visco-elastic forces in the PDEs (\ref{eq:Fep}) and (\ref{eq:Fve}) by integrating them over small time increments $dt \ll \xi/k_2$ during the entire duration of the test. This was performed by using a forward finite difference discretisation of the time derivatives.

We tested the ability of this model to capture the drag force throughout relaxation tests by
\begin{itemize}
\item Measuring the peak drag force $F_0$; 
\item Measuring the slider force as $\mathcal{F}_s = F(t\to \infty)$;
\item Defining the viscous parameter as $\xi = \frac{F_0-\mathcal{F}_s}{v}$;
\item Estimating the stiffness $k_{2}$ by fitting the drag force curve $F(P)$ during the relaxation  ($t\geqslant t_0$) by Eq. (\ref{eq:relax}, using $\lambda = \xi/k_2$ and $k_2$ as sole free fitting parameter. 
\item Estimating the stiffness $k_{1}$ by fitting the drag force curve $F(P)$ at low deformation by ($P\ll P_0$) by the function $F= (k_1+k_2) P$, using $k_1$ as sole free fitting parameter. 
\end{itemize}

Figure \ref{fig10} compares the response of this analogue to the measured drag force for relaxation tests performed with three different initial velocities. These results suggest that the analogue can capture the salient properties of the drag force, including the pre-peak rise and peak values, as well as its time of relaxation.
The analogue parameters used that best match the measurements are summarised on Table \ref{tab:model_para}. It appears that the values of the stiffnesses $k_{1,2}$, the viscous parameter $\xi$ and the slider force $\mathcal{F}_s$  depend on the test velocity. Specifically, the stiffness $k_1$ and the viscous parameter becomes significantly lower at high velocities. Coincidently, the slider force $\mathcal{F}_s = F(t\to \infty)$ decreases at high velocity, and is always significantly lower than the quasi-static peak drag  measured for this plate ($F_0^{qs} =$ \SI{10.3}{N}).

A possible explanation for these effects would be that the high-pressure gradient developing in the pores at high velocity could partially mobilise the granular matrix: some grains could move and possibly become fluidised under the action of the liquid pressure gradient. This could induce a rate-softening and rate-weakening of the granular matrix explaining the reduction in stiffnesses and effective strength $\mathcal{F}_s$ at high velocities. However, the results presented here do not provide direct evidence of such a process, which therefore remains conjectural.

\begin{table}[!htb]

\begin{center}
\begin{tabular}{c | c | c | c | c | c | c}
%\hline
 $v$ & $F_0$ & $\mathcal{F}_s$ & $k_1$ & $\xi$ & $\lambda$ & $k_2$ \\
mm/min   &N & N & $\times 10^3$ N/m &$\times 10^3$ N/(m/s) &s &  $\times 10^4$ N/m \\
\hline
 $10$      &  10.5    & 8.5  & $46$  &  $10$             & $0.29$    & $3.4$ \\  \hline
 $50$      &   10.3  & 6.3 & $46$  &  $4.5$             & $0.27$    & $1.6$ \\  \hline
 $100$   &  10.0   & 6.3 & $41$  &  $3.4$             & $0.25$    & $1.4$ \\  \hline
  $200$   &   22.3  & 6.2 & $21$  &  $5.0$             & $0.19$    & $2.6$ \\  \hline
 $400$     &  25.0    & 5.8    & $1.3$  &  $3$ & $0.13$    & $2.4$ \\  \hline
 $600$     &   32.0   &   4.9 & $1.3$  &  $3.5$ & $0.13$    & $2.3$ \\  \hline
 $800$     &  36.1   &  4.1   & $1.2$  &  $3$ & $0.13$    & $2.5$ \\  \hline
$1\,000$ &  47.6 & $4.1$    & $1.3$  &  $3$ & $0.11$  & $2.8$ \\  
\end{tabular}
\caption{Parameters used to test the visco-elasto-plastic model in figure \ref{fig9} against experimental relaxation tests performed with $B=$\SI{40}{mm} and $d$ = \SI{0.3}{mm} at three different velocities. For all tests, the quasi-static peak drag force was:  $F_0^{qs}$ = \SI{10.3}{N}. The peak force $F_0$ was measured from experiments, and the value of the viscous parameter $\xi$ was deduced accordingly; both stiffnesses $k_{1,2}$ were inferred by fitting experimental data  (see text). 
 \label{tab:model_para} }
\end{center}
\end{table}%

\subsection{Post-peak drag force drop}

The visco-elasto-plastic analogue can reproduce the experimental drag force evolution in relaxation tests when the velocity is stopped shortly after the peak drag force is reached. However, it does not include a mechanism that lowers the drag force when the velocity is kept constant. It instead predicts that the drag force would reach a maximum $\mathcal{F}_s+\xi v$ and then plateau. This is not consistent with the measurements shown in figure \ref{fig2}, which evidence a strong decrease in drag force after the peak is reached ($P>P_0$). In comparison with the peak drag force, the post-peak drag force only marginally increases at high velocities. This suggests that the viscous component of the drag force becomes much weaker after the peak. 

We hypothesise that this effect may be caused by the circulation of grains around the plate taking place at large displacements. Such granular flows around moving objects have been consistently observed in a variety of mobility tests in dry conditions \cite{candelier2009creep,harich2011intruder,kolb2013rigid,seguin2019hysteresis,takada2020drag,allen2019effective}.  
Our hypothesis is that grains do not significantly recirculate before the peak drag force is reached, which is consistent with the Darcy-flow mechanism we proposed to explain the peak drag force. We further hypothesise that grains could start circulating around the plate at large displacements, thus moving along with the liquid. Having the granular matrix moving with the liquid would significantly reduce the magnitude of the viscous forces, resulting in a drop in drag force. 
Ref. \cite{allen2019effective} provides a detailed analysis of the post-peak drag forces measured in horizontally dragged rods. The study points out that this force may be explained by the frictional stresses developing in the shear granular packing. This is consistent with our measurements of a vanishing drag force when the plate reaches the free surface, and the normal stress in the granular packing around it vanishes.

%\section{Mechanical analogue}

%\begin{figure*}[h!]
 %{\includegraphics[width=1\textwidth]{figure/model_relax.eps}}
%\caption{The result of $1-D$ simulation to validate the proposed viscoplastic model is shown where (a) the anchor is displaced to $1.5$ $\sim$ $4.2$ $mm$ at uplift speeds $10$, $400$ and $1000$ $mm/min$ and kept at that position for $20$ sec. The forces due to solid and viscous components are shown separately as well as combining together, which is also super-imposed by original experimental data at uplift speeds (b) 10 mm/min (c) 400 mm/min and (d) 1000 mm/min.}
%\label{model_relax}      
%\end{figure*}

\section{Conclusion}

This study pointed out that the presence of water in the pores can quantitatively and qualitatively modify granular drag forces. 

By focusing on an elementary drag test, we found that water induces a viscous component, which adds to the frictional drag developing in dry conditions. 
This introduces a rate-dependence for the drag force, which increases with the plate velocity. Specifically, we found a linear relationship for the peak drag force $F_0 (v)$ (see Eq. \ref{eq:F_v}). We introduced a \textit{Darcy-flow} mechanism, which rationalised the dependence of $F_0$ with the plate velocity, the grain size and the plate size. The resulting model is expressed in Eqs. (\ref{eq:F0decomposition}), (\ref{eq:F0qs}) and (\ref{eq:Fvis}).\\
%/Users/prognon/Dropbox (Sydney Uni)/PhD/tanvir/Wet_drag/wet_drag.bbl

We further found that this peak drag force gradually relaxes in time when the plate's motion is stopped and showed how this behaviour is analogous to a visco-elasto-plastic dynamics. This observation suggests that peak drag force $F_0$ measured in constant velocity tests does not correspond to a static resistance. According to the elasto-visco-plastic analogue we introduced, a plate loaded with a constant external force greater than $F_0^{qs}$ would not significantly move during a period of time and then would start moving through the packing. While such delays in deformation response have been reported with immersed granular materials subjected to step shear-stress changes \cite{cassar2005submarine,rognon2010internal,rognon2011flowing}, they remain to be experimentally observed in drag experiments.

Our results also evidenced a significant drop in drag force at large displacements after the peak is reached. In comparison with peak drag forces, post-peak drag forces appear to have a much weaker rate-dependence. 
We conjectured that this could result from the development of grain recirculation around the plate, which would reduce the relative velocity between the liquid and grains and thus reduce the magnitude of the viscous forces. In-situ visualisation techniques \cite{baker2018x} or particle-based simulations could possibly provide a direct evidence of this mechanism.   

Lastly, we expect these water-induced rate-effects to play an important role in the mobility response of objects subjected to dynamic and cyclic loadings \cite{athani2018mobility,athani2019inertial}, and in the development of drag instabilities that have been observed in dry packings \cite{gravish2010force,duri2017vertical,hossain2020rate}.

%\bibliography{biblio}   % name your BibTeX data base

\begin{thebibliography}{10}
\providecommand{\url}[1]{{#1}}
\providecommand{\urlprefix}{URL }
\expandafter\ifx\csname urlstyle\endcsname\relax
  \providecommand{\doi}[1]{DOI \discretionary{}{}{}#1}\else
  \providecommand{\doi}{DOI \discretionary{}{}{}\begingroup
  \urlstyle{rm}\Url}\fi

\bibitem{andreotti2013granular}
B.~Andreotti, Y.~Forterre, O.~Pouliquen, \emph{Granular media: between fluid
  and solid} (Cambridge University Press, 2013)

\bibitem{albert1999slow}
R.~Albert, M.~Pfeifer, A.L. Barab{\'a}si, P.~Schiffer, Physical review letters
  \textbf{82}(1), 205 (1999)

\bibitem{albert2000jamming}
I.~Albert, P.~Tegzes, B.~Kahng, R.~Albert, J.~Sample, M.~Pfeifer, A.L.
  Barabasi, T.~Vicsek, P.~Schiffer, Physical review letters \textbf{84}(22),
  5122 (2000)

\bibitem{albert2001granular}
I.~Albert, J.~Sample, A.~Morss, S.~Rajagopalan, A.L. Barab{\'a}si, P.~Schiffer,
  Physical Review E \textbf{64}(6), 061303 (2001)

\bibitem{miller1996stress}
B.~Miller, C.~O'Hern, R.~Behringer, Physical Review Letters \textbf{77}(15),
  3110 (1996)

\bibitem{nguyen1999properties}
M.~Nguyen, S.~Coppersmith, Physical Review E \textbf{59}(5), 5870 (1999)

\bibitem{gravish2010force}
N.~Gravish, P.B. Umbanhowar, D.I. Goldman, Physical review letters
  \textbf{105}(12), 128301 (2010)

\bibitem{costantino2011low}
D.~Costantino, J.~Bartell, K.~Scheidler, P.~Schiffer, Physical Review E
  \textbf{83}(1), 011305 (2011)

\bibitem{ding2011drag}
Y.~Ding, N.~Gravish, D.I. Goldman, Physical Review Letters \textbf{106}(2),
  028001 (2011)

\bibitem{guillard2013depth}
F.~Guillard, Y.~Forterre, O.~Pouliquen, Physical review letters
  \textbf{110}(13), 138303 (2013)

\bibitem{takada2020drag}
S.~Takada, H.~Hayakawa, Granular Matter \textbf{22}(1), 6 (2020)

\bibitem{percier2011lift}
B.~Percier, S.~Manneville, J.N. McElwaine, S.W. Morris, N.~Taberlet, Physical
  Review E \textbf{84}(5), 051302 (2011)

\bibitem{potiguar2013lift}
F.Q. Potiguar, Y.~Ding, Physical Review E \textbf{88}(1), 012204 (2013)

\bibitem{takehara2010high}
Y.~Takehara, S.~Fujimoto, K.~Okumura, EPL (Europhysics Letters) \textbf{92}(4),
  44003 (2010)

\bibitem{takehara2014high}
Y.~Takehara, K.~Okumura, Physical review letters \textbf{112}(14), 148001
  (2014)

\bibitem{cassar2005submarine}
C.~Cassar, M.~Nicolas, O.~Pouliquen, Physics of fluids \textbf{17}(10), 103301
  (2005)

\bibitem{rognon2010internal}
P.~Rognon, I.~Einav, C.~Gay, Physical Review E \textbf{81}(6), 061304 (2010)

\bibitem{rognon2011flowing}
P.G. Rognon, I.~Einav, C.~Gay, Journal of Fluid Mechanics \textbf{689}, 75
  (2011)

\bibitem{degiuli2015unified}
E.~DeGiuli, G.~D{\"u}ring, E.~Lerner, M.~Wyart, Physical Review E
  \textbf{91}(6), 062206 (2015)

\bibitem{ikeda2019universal}
A.~Ikeda, T.~Kawasaki, L.~Berthier, K.~Saitoh, T.~Hatano, arXiv preprint
  arXiv:1904.07359  (2019)

\bibitem{allen2019effective}
B.~Allen, A.~Kudrolli, Physical Review E \textbf{100}(2), 022901 (2019)

\bibitem{dyson2014pull}
A.~Dyson, P.~Rognon, G{\'e}otechnique Letters \textbf{4}(4), 301 (2014)

\bibitem{hossain2020rate}
T.~Hossain, P.~Rognon, arXiv preprint arXiv:2001.07880  (2020)

\bibitem{das2013earth}
B.M. Das, S.K. Shukla, \emph{Earth anchors} (J. Ross Publishing, 2013)

\bibitem{meyerhof1968ultimate}
G.~Meyerhof, J.~Adams, Canadian geotechnical journal \textbf{5}(4), 225 (1968)

\bibitem{rowe_behaviour_1982}
R.K. Rowe, E.H. Davis, Geotechnique \textbf{32}(1), 25 (1982)

\bibitem{murray_uplift_1987}
E.J. Murray, J.D. Geddes, Journal of Geotechnical Engineering \textbf{113}(3),
  202 (1987)

\bibitem{merifield2006ultimate}
R.~Merifield, S.~Sloan, Canadian Geotechnical Journal \textbf{43}(8), 852
  (2006)

\bibitem{kumar2008vertical}
J.~Kumar, K.~Kouzer, Canadian Geotechnical Journal \textbf{45}(5), 698 (2008)

\bibitem{khatri2011effect}
V.N. Khatri, J.~Kumar, Canadian Geotechnical Journal \textbf{48}(3), 511 (2011)

\bibitem{bhattacharya2014pullout}
P.~Bhattacharya, J.~Kumar, Canadian Geotechnical Journal \textbf{51}(11), 1365
  (2014)

\bibitem{askari2016intrusion}
H.~Askari, K.~Kamrin, Nature materials \textbf{15}(12), 1274 (2016)

\bibitem{giampa2018effect}
J.~Giampa, A.~Bradshaw, H.~Gerkus, R.~Gilbert, K.~Gavin, V.~Sivakumar,
  G{\'e}otechnique pp. 1--9 (2018)

\bibitem{sakai1998particle}
T.~Sakai, T.~Tanaka, Soils and Foundations \textbf{38}(2), 93 (1998)

\bibitem{athani2017grain}
S.~Athani, P.~Kharel, D.~Airey, P.~Rognon, G{\'e}otechnique Letters pp. 1--7
  (2017)

\bibitem{costantino_starting_2008}
D.J. Costantino, T.J. Scheidemantel, M.B. Stone, C.~Conger, K.~Klein, M.~Lohr,
  Z.~Modig, P.~Schiffer, Physical Review Letters \textbf{101}(10) (2008)

\bibitem{athani2019inertial}
S.~Athani, P.~Rognon, Physical Review Fluids \textbf{4}(12), 124302 (2019)

\bibitem{carman1939permeability}
P.C. Carman, The Journal of Agricultural Science \textbf{29}(2), 262 (1939)

\bibitem{rognon2014explaining}
P.~Rognon, M.~Macaulay, D.~Griffani, I.~Einav, EPL (Europhysics Letters)
  \textbf{108}(3), 34004 (2014)

\bibitem{candelier2009creep}
R.~Candelier, O.~Dauchot, Physical review letters \textbf{103}(12), 128001
  (2009)

\bibitem{harich2011intruder}
R.~Harich, T.~Darnige, E.~Kolb, E.~Cl{\'e}ment, EPL (Europhysics Letters)
  \textbf{96}(5), 54003 (2011)

\bibitem{kolb2013rigid}
E.~Kolb, P.~Cixous, N.~Gaudouen, T.~Darnige, Physical Review E \textbf{87}(3),
  032207 (2013)

\bibitem{seguin2019hysteresis}
A.~Seguin, The European Physical Journal E \textbf{42}(1), 13 (2019)

\bibitem{baker2018x}
J.~Baker, F.~Guillard, B.~Marks, I.~Einav, Nature communications \textbf{9}(1),
  1 (2018)

\bibitem{athani2018mobility}
S.~Athani, P.~Rognon, Granular Matter \textbf{20}(4), 67 (2018)

\bibitem{duri2017vertical}
A.~Duri, S.~Mandato, F.~Mabille, B.~Cuq, T.~Ruiz, in \emph{EPJ Web of
  Conferences}, vol. 140 (EDP Sciences, 2017), vol. 140, p. 03083

\end{thebibliography}

 \end{document}